\documentstyle[12pt]{article}
\newcommand\qovm{\frac{q}{2M}}

\newcommand\mnuq{\frac{m_{\nu}}{q}}
\def\bsig{\hbox{\boldmath $\sigma$}} 
\def\muar{\buildrel{\mu^-}\over\longrightarrow}
\def\movq{\frac{m_{\nu}^2}{4q^2}}
\def\mvdq{{1 \over 2}{\frac{m_{\nu}^2}{q^2}}}
\newcommand{\shel}{{\sum\limits_h}}
\newcommand{\CG}[6]{\left(\begin{array}{cc|c} \!#1\! & \!#3\! & 
\!#5\! \\%
            \!#2\! & \!#4\! & \!#6\!\\  \end{array} \right) }
\newcommand{\Cma}{\CG{J_i}{M_i}{L}{M}{J_f}{M_f}}  
\newcommand\fhJf{\frac{\sqrt{2}}{{\hat J}_f}}
\newcommand{\Pm}{{P_{\mu}}}  
\newcommand\bPm{{\bf P}_{\mu}}  
\newcommand{\bJf}{{{\bf J}_f}}  
\newcommand{\bPf}{{{\bf P}_f}}  
\newcommand\eif{\exp{[i\phi_h]}}
\newcommand\hr{\hat{\bf r}}
\newcommand\hq{\hat{\bf q}}
\newcommand\hz{\hat{\bf z}}
\newcommand\hL{\hat L}

\def\sqF{\frac{\alpha Z}{2}}

\def\pM{i\frac{\bf p}{M}}
\newcommand{\bear}{\begin{eqnarray}}
\newcommand{\eear}{\end{eqnarray}}
\begin{document}
\medskip
\vskip 2.0 cm
\begin{center}
{\Huge Relativistic and  Neutrino Mass Effects\\ in Partial Muon 
Capture \footnote{Submitted for publication in Journal Phys. G }\\}
\vskip 1.2 cm
\begin{em}
\begin{Large}
by\\Stanis{\l}aw Ciechanowicz \footnote{e-mail: 
ciechano@proton.ift.uni.wroc.pl}\\{\em Institute of Theoretical 
Physics, Wroc{\l}aw University\\PL 50-204 Wroc{\l}aw, 
pl. Maksa Borna 9, Poland}\\
\vskip 0.4 cm
Zbigniew Oziewicz \footnote{A member of Mexican Sistema Nacional de 
Investigadores\\e-mail: oziewicz@servidor.unam.mx}\\{\em Facultad de 
Estudios 
Superiores, Cuautitl\'{a}n, UNAM\\
Apartado Postal \# 25, C.P. 54700, Estado de M\'{e}xico}\\
\vskip 0.4 cm
Nicolai Popov \footnote{e-mail: popov@cip.physik.uni-
muenchen.de}\\{\em Sektion 
Physik der Universit\"at M\"unchen\\
Schellingstr. 4, 80799 M\"unchen, Germany}\\
\end{Large}
\end{em}
\end{center}
\vskip 1.8 cm
\begin{abstract}
\baselineskip15pt
\begin{bf}
The characteristics of the partial nuclear muon capture with
massive left-handed Dirac neutrino and relativistic component of
the muon wave function have been derived. The multipole
amplitudes are given as a function of neutrino mass parameter and
reduced nuclear matrix elements which are modified by the small 
component 
of the muon wave function. As an example, the capture rate,  
asymmetry  and 
polarization of recoil nuclei are investigated in terms of these 
multipole 
amplitudes.
\baselineskip15pt
\end{bf}
\end{abstract}
\vfill
\hfill {\Large \bf March 1996 ITP UWr 906/96}
\thispagestyle{empty}

\newpage
\section{Introduction}
\medskip
\par
Neutrino mass problem is a topic of many theoretical and
experimental works. In recent years a considerable 
effort has been devoted to find massive neutrino in the
different processes. It has been extensively looked for in
neutrino oscillations, beta decay, double beta decay $\beta \beta$ 
(especially neutrinoless beta decay $\beta \beta (0 \nu))$, charged 
lepton 
conversion as well as in cosmology and astrophysics (see e.g. [1-
4]).\\
Massive neutrinos are naturally predicted beyond the standard model 
of 
electroweak interaction, e.g. by "{\em see-saw}" mechanism. 
One can imagine also the gravitationally induced neutrino masses 
[5].\\ 
Experimental data of $\beta \beta$ and $\beta \beta(0\nu)$ decay give 
limits 
on the mass of Majorana neutrino and right-handed currents [6]. The 
solar 
neutrino puzzle can be explained in frame of {\em MSW} mechanism 
[7,8], or 
by long-wavelength vacuum oscillations [9].\\
As for muon capture process, the possibility to verify the existence 
of  
massive muon neutrino was proposed in ref. [10]. The calculations 
of  nuclear muon capture observables with massive neutrino 
were undertaken in refs. [11,12]. So far, however, the calculations 
were 
performed without the small relativistic corrections which {\em a 
priori} 
could be competed with neutrino mass effect. We present here the 
results of 
our calculations of the observables of  nuclear muon capture, taking 
into 
account both the neutrino mass effect and relativistic component of 
the muon 
wave function.
\section{Effective Hamiltonian and multipole expansion}
As earlier, we use the method of multipole expansion [13, 14] for the
effective Hamiltonian in the nuclear muon capture on the
lepton-hadron level, i.e. Fermi theory of four-fermion point
interaction. The calculation is performed with relativistic muon and 
massive
left-handed Dirac neutrino. 

We take the effective $\mu $-capture Hamiltonian in the standard 
current $\times $ current form
\bear
 <{\bf q}, {1\over 2} \nu ,\hz 
\mid H_{\mu} \mid {1\over 2} m,\hz > = GJ_{\lambda}l_{\lambda 
},\qquad  
\eear
\noindent where 
$G = \frac {G_{\mu }\cos \theta _{c}}{ \sqrt{2}}$
is the coupling constant, $J_{\lambda }$ is the hadron current 
operator, {\bf q} is  vector of neutrino momentum, and
\bear
l_{\lambda }~=~\overline{\Psi}_{\nu _{\mu }}\gamma _{\lambda 
}(1~+~ \gamma _{5})\Psi _{\mu }\qquad  
\eear
\noindent is matrix element of the modified lepton current with the 
{\it massive left-handed Dirac neutrino}. With $\hz $, unit
vector along $z $ axis, we denote the direction for lepton spin
quantization. 

The Dirac bispinor is commonly used for the neutrino wave function,
\bear
\Psi _{\nu _{\mu }} \equiv  <{\bf r}\mid {\bf q}, {1\over 2} \nu 
,\hz> 
= {\frac{1} {\sqrt{2}}} \exp \{+\hbox{i{\bf q}}\cdot {\bf r\}}
\left( \matrix{{-c_{1}(\bsig  \cdot \hq)} \cr {c_{0}} \cr} \right)
\otimes \chi _{\nu },\qquad 
\eear 
where $c_{0} $ and $c_{1} $ are energy-mass coefficients,
\bear
c_{o} = \sqrt{1 + {m_{\nu} \over E_{\nu}}},\qquad 
c_{1} = \sqrt{1 - {m_{\nu} \over E_{\nu}}}.\qquad 
\eear
We consider the nuclear partial transition in the muon capture 
process;
\bear
\mu^- + (A, Z)_{J_i} \rightarrow (A, Z - 1)_{J_f} + \nu_{\mu}.
\eear
With the method of [13], we get the multipole expansion;
\bear
<-{\bf q},J_{f}M_{f};{\bf q} {1\over 2} h,\hq \mid H_{\mu }\mid
J_{i} M_{i}; {1\over 2} \mu ,\hq> \equiv \sum_{LM} i^{-L} \hL 
{\cal T}^{h\mu }_{L} \Cma D^{L}_{M\mu -h}(\Omega_{\hq}). 
\eear 
The multipole amplitudes 
${\cal T}^{h\mu }_{L} $ ($h, \mu  = \pm  {1\over 2} $, 
$\mid J_i - J_f \mid \leq L \leq J_i + J_f $ ) 
are  given  by linear combinations of the coupling constants and the
reduced matrix elements in the nuclear, or in general, hadron space.
With $\hq $, we denote that lepton spin is quantized now along the
direction of neutrino momentum.

We find, that neutrino energy-mass coefficients factorize from the
multipole amplitudes ${\cal T}^{h\mu }_{L}$ as follows;
\bear
{\cal T}^{h\mu }_{L} = \left(\frac{c_{0}- 2hc_{1}}{2}\right)T^{h\mu 
}_{L}.\qquad 
\eear
It should be noted that still ${\cal T}^{h\mu }_{L}$ are matrix 
elements of 
lepton current, where the diagonal terms ${\cal T}^{hh}_{L}$ are 
amplitudes 
with no spin-flip of the lepton  in the weak current, while 
${\cal T}^{h-h}_{L}$ are the amplitudes where the lepton spin is 
flipped in course of interaction.  These amplitudes essentially differ
for the neutrino helicities $h = +{1\over 2}$ and $h~= -{1\over 2}$. 
We
have, like in [13], the following expressions for the multipole
amplitudes $T^{h\mu }_{L}$ in terms of the reduced matrix elements in
hadron space;
\bear
\lefteqn{T^{hh}_{L}  = } \\
&& [0LL,J_4]_{hh} + i[1LL,{\bf J}]_{hh} + \root\of{\frac{L}{2L+1}}
[1L-1L,{\bf J}]_{hh} + 
\root\of{\frac{L+1}{2L+1}}[1L+1L,{\bf J}]_{hh}, \qquad
\nonumber \eear 
\bear
\lefteqn{T^{h-h}_{L}  = } \\
&& [0LL,J_{4}]_{h-h} - i[1LL,{\bf J}]_{h-h} + 
\root\of{\frac{L+1}{2L+1}} [1L-1L,{\bf J}]_{h-h} +
\root\of{\frac{L}{2L+1}}[1L+1L,{\bf J}]_{h-h}.\qquad
\nonumber \eear 
Here, $[0LL,J_{4}]_{h\pm h}$ and $[1L'L,{\bf J}]_{h\pm h}$ are
notations for the reduced nuclear matrix elements, in the sense of the
Wigner-Eckart theorem.

For the relativistic muon in the $1s $ atomic state ($K-shell $), we
take the Dirac bispinor for the particle bound in the radial
electrostatic potential;
\bear
\Psi_{\mu } &= &
\left( \matrix{{-iF(r)\chi_{1m}(\hr)} \cr {G(r)\chi_{-1m}(\hr)} \cr}
\right),\qquad 
\eear
with the customary definition of spherical spinors [15],
\bear
\chi _{km}(\hr) = \left[Y^{l}(\hr) \otimes \chi_{1/2}\right]_{Jm}, 
\qquad 
(J~=~l-{1\over2} \hbox{sign}k). \qquad 
\eear
$G(r)$ and $F(r)$ are named as large and small radial components,
respectively.

Next, these matrix elements are given explicitly in the product form 
of
two terms, where the former is bilinear combination of the radial
functions which emerge from the matrix elements of the leptonic 
current
due to the neutrino plane wave and $1s$ muon bound state, and the
latter is tensor operator that in the standard way couples the 
neutrino
spherical harmonics and hadron current operators:
\bear
\lefteqn{[0LL,J_{4}]_{hh}  = } \\
&& \fhJf <f\parallel G \{G(r)j_{L}(qr) - 2hF(r)[\frac{L}{2L+1} j_{L-
1}(qr) - 
\frac{L+1}{2L+1}j_{L+1}(qr)]\} J_{4} Y^{L}(\hr) \parallel i>, 
\nonumber \eear 
\bear
\lefteqn{[0LL,J_{4}]_{h-h}  = } \\ 
&& \fhJf <f \parallel G  (-2h)F(r) {\sqrt{L(L+1)} \over {2L+1}}
[j_{L-1}(qr) +j_{L+1}(qr)] J_{4} Y^{L}(\hr) \parallel i>, 
\nonumber \eear 
\bear
\lefteqn{[1LL,{\bf J}]_{hh}  = } \\
&& \fhJf <f \parallel G F(r)
\frac{\sqrt{L(L+1)}}{2L+1} [j_{L-1}(qr) +j_{L+1}(qr)]
[{\bf J} \otimes  Y^{L}(\hr)]^{L} \parallel i>, 
\nonumber \eear 
\bear 
\lefteqn{[1LL,{\bf J}]_{h-h}  = } \\
&& \fhJf <f \parallel G  \{G(r) j_{L}(qr) - 
F(r) [{\frac{L+1}{2L+1}}j_{L-1}(qr) - {\frac{L}{2L+1}}j_{L+1}(qr)]\}
[{\bf J} \otimes Y^{L}(\hr)]^{L} \parallel i>, 
\nonumber \eear
\bear
\lefteqn{[1L-1L,{\bf J}]_{hh}  = } \\
&& \fhJf <f \parallel G (-2h) [G(r)j_{L-1}(qr)
+ 2h F(r) j_{L}(qr)] [{\bf J} \otimes Y^{L-1}(\hr)]^{L} \parallel i>, 
\nonumber \eear
\bear 
\lefteqn{[1L-1L,{\bf J}]_{h-h} = } \\
&& {\fhJf <f \parallel
G  (-2h) [G(r) j_{L-1}(qr) + F(r) j_{L}(qr)] [{\bf J} \otimes
Y^{L-1}(\hr)]^{L} \parallel i> },
\nonumber \eear 
\bear 
\lefteqn{[1L+1L,{\bf J}]_{hh}  = } \\
&&{\fhJf <f \parallel G  (-2h) [G(r) j_{L+1}(qr)
- 2h F(r) j_{L}(qr)] [{\bf J} \otimes Y^{L+1}(\hr)]^{L} \parallel i> 
},
\nonumber \eear
\bear 
[\lefteqn{1L+1L,{\bf J}]_{h-h}  = } \\
&& {\fhJf <f \parallel
G  (-2h) [-G(r) j_{L+1}(qr) + F(r) j_{L}(qr)] [{\bf J} \otimes
Y^{L+1}(\hr)]^{L} \parallel i> }.
\nonumber \eear
\noindent $j_{v}(qr)$ are the neutrino spherical Bessel functions.
Tensor couplings of the hadron current spatial components ${\bf J}$ 
and
spherical harmonics from the neutrino plane wave are as follows,
\bear
[ {\bf J} \otimes Y^{L'}(\hr)]^{L}_{M} & = & \sum_{\mu m}
 \CG{1}{\mu}{L'}{m}{L}{M}
({\bf J})^{1}_{\mu} Y^{L'}_{m}(\hr).\qquad 
\eear 
\section{Relativistic muon and massive neutrino}
Question now is for the multipole amplitudes to compare the leading
relativistic muon (RLMU) effects with bare massive neutrino (NEMA)
contribution.  To answer this, at least qualitatively in $0^{\pi_i}
\muar 1^{\pi_f} $ transitions, we have included RLMU effects into the
formulae for multipole amplitudes calculated within the framework of
the impulse approximation for the hadron current.  Then, we generalize
Fujii-Primakoff form factors to;
\bear
G^h_{V} &=& g_{V}\pmatrix{1 - 2h \qovm }, \nonumber\\
G^h_{A} &=& g_{A} + 2h g_{V}(1 + \mu_{p} - \mu_{n}) \qovm, \\
G^h_{A} - G^h_{P} &=& -2h g_{A} + \pmatrix{g_{A} - g_{P}
+ 2 \delta_{{1 \over 2}h} g_{P}{q \over m_{\mu}}} \qovm.
\nonumber \eear
It is convenient for our present purposes to use point nucleus 
solution
(Coulomb potential). Then the well known relation holds;
\bear
F(r) &= & - \sqrt{\frac{1 - \gamma}{1 + \gamma}}G(r) 
\simeq -\frac{\alpha Z}{2}G(r),
\eear
where $ \gamma = \sqrt{1 - (\alpha Z)^2} $. $\alpha $ is a fine 
structure 
constant and $Z $ - atomic number. We have considered contributions 
due to 
$F(r) $ together with derivative $G'(r) $.

Reduced matrix elements are written within the short notation;
\bear
 \{1L'L,j_l\bsig\} \equiv \fhJf < f \Vert \sum\limits_{i = 1}^A
G(r_i) j_{l}(qr_i) [{\bsig_i} \otimes  Y^{L'}(\hr_i)]^{L} \tau_i^-
\Vert i > 
\eear
and additionally 
$\{1L'L,j_{L'}\bsig \} \equiv [1L'L,\bsig] $, 
if the labels in Bessel function and spherical harmonics are the 
same.\\
Below, the multipole amplitudes in $0^{\pi_i} \muar 1^{\pi_f} $
transitions with the leading order RLMU contributions, $\sim \sqF $,
are given;
\bear
 iT^{hh}_1 ({\scriptstyle \pi_i=\pi_f}) &= & (G^h_{A} -
G^h_{P}) \sqrt{1 \over 3} \left( [101,\bsig ] + \sqrt{2}
[121,\bsig ] \right) + g_A [011,\bsig \cdot \pM ] \\
&&\null + \sqF g_A \sqrt{1 \over 3} \left( \{ 101,j_1 \bsig \} -
\sqrt{2} \{121,j_1 \bsig \} \right),
\nonumber \eear
\bear
iT^{h-h}_1 ({\scriptstyle \pi_i=\pi_f}) &= &
 - 2h G^h_{A}\left( [101,\bsig ] - \sqrt{2} [121,\bsig ] \right) 
- g_V [111, \pM ] \\
&&\null - 2h  \sqF g_A \left( \{ 101,j_1 \bsig \} - \sqrt{2} 
\{ 121,j_1 \bsig \} \right),
\nonumber \eear
\bear
 T^{hh}_1 ({\scriptstyle \pi_i=-\pi_f}) &= & G^h_{V} [011,{\bf 1 }] 
+ 2h g_{V} \sqrt{1 \over 3} \left( [101,\pM] + \sqrt{2}
[121,\pM] \right) 
\\
&&\null + \sqF \cdot \frac{1}{3} \left( 2h g_V \{ 011,j_{0}{\bf1} \} -
 g_A
\sqrt{2} \{ 111,j_{0} \bsig \} \right),
\nonumber \eear
\bear
 T^{h-h}_1 ({\scriptstyle \pi_i=-\pi_f}) &= &
 -G^h_{A} [111,\bsig ]
+ 2h g_V \left( [101,\pM] - \sqrt{2} [121,\pM] \right) 
\\
&&\null + \sqF \left( 2hg_V \frac{\sqrt{2}}{3} \{011,j_{0} {\bf 1} \}
+ g_A 2 \{111,j_{0} \bsig \} \right).
\nonumber \eear

The largest RLMU contributions are given here.  These are only from 
the
small radial component of the muon function $F(r) $.  All reduced
matrix elements with Bessel functions of rank $L-2 $ and $L-3 $ have
been omitted as they can not contribute for multipolarity $L=1 $.

Estimating NEMA contributions we perform with Lommel integral for
spherical Bessel functions for ($q < E $) in case of {\em massive },
and ($q = E $) {\em massless } neutrinos;
\bear
I_v(q) = \int\limits_0^{\infty} j_v(qr) dr = \frac{\sqrt{\pi}}{2} 
\cdot \frac{\Gamma({{v+1} \over 2})}{\Gamma({{v+2} \over 2})}
\cdot {1 \over q}
\eear
With this, the difference between $j_v(qr) $ and $j_v(Er) $ in radial
integral with nuclear wave functions can be evaluated;
\bear
I_v(q) = {1 \over {c_0 c_1}} I_v(E) \simeq \left(1 + \mvdq \right) 
I_v(E)
\eear
Multipole amplitudes $T^{h \mu}_L $ change approximately by a global
factor $(1 + \mvdq) $. We take the upper limit $ m_{\nu}c^2 \leq
160~keV $ from [16], and $q \simeq 100 MeV $.  The magnitude of 
$\mvdq 
= 1.28 \times 10^{-6} $ measures the change of the amplitudes. 
Coulomb coupling parameter $\alpha Z $ for e.g. $^{12}C $ gives
${1 \over 2} \alpha Z \simeq 2.16 \times 10^{-2} $. 
This suggests RLMU effect to the multipole amplitudes by four orders 
of magnitude stronger than NEMA contribution.  
\section{Observables in muon capture}
        In order to obtain the capture rate and correlation
characteristics in muon capture, we use the polarization density
matrix for the final state of the system, 
\bear
\lefteqn{<-{\bf q}, J_f M_f; {\bf q}, \frac{1}{2} \nu \mid\rho_f
\mid -{\bf q}, J_f M_f^{'}; {\bf q}, \frac{1}{2} \nu^{'} >  = } \\ 
& = &  \frac{1}{2(2J_{i} + 1)} \sum_{M_i,m,m^{'}}  
<-{\bf q}, J_f M_f; {\bf q}, \frac{1}{2} \nu \mid H_{\mu} 
\mid J_i M_i;\frac{1}{2} m >
<\frac{1}{2} m \mid 1 + 2 \bPm \cdot {\bf s}_{\mu} \mid \frac{1}{2} 
m^{'}> 
\nonumber \\
& & < -{\bf q}, J_{f} M_{f}^{'}; {\bf q}, \frac{1}{2}\nu^{'} 
\mid H_{\mu} \mid J_i M_i; \frac{1}{2} m^{'} >^*,  
\nonumber \eear 
where direction of $\hz$ is the spin quantization axis for both 
nucleus
and leptons.  $\bPm $ is the residual muon polarization on the K-orbit
and ${\bf s}_{\mu} = {1 \over 2 } \bsig $ denotes the muon spin
operator.\\
Averaging over the initial states and summing over the final ones,
which are not measured in the experiment, we obtain the formulae for
nuclear muon capture rate, 
$\Lambda_c \sim  Tr \overline{\rho}_{f}
\equiv {1\over 4\pi}\ \! \int \! d{\hat \bPm} {1\over 4\pi} \!
\int d\hq \, Tr\rho_{f} $, 
recoil polarization vector 
$ \bPf \equiv Tr\left(\bJf \rho_{f}\right) \big/ J_f Tr 
\overline{\rho}_{f} $, 
and the recoil asymmetry 
$ {\cal W}(\theta) \equiv Tr\rho_{f} \big/  Tr \overline{\rho}_{f} 
$.\\
The formula for muon capture rate is; 
\bear
\Lambda_c =  c_0 c_1 N_{fi} \frac{2J_f + 1}{2(2J_i + 1)} 
\shel \left( {{1 - 2hc_{0}c_{1}} \over 2} \right) \lambda_h,
\eear
here $c_0 c_1 N_{fi} $ is a phase space volume, with conventional
factor for $m_{\nu} = 0 $;
\bear
N_{fi} = \frac{1}{2 \pi}  q^2 \left[ 1 + \frac{m_{\mu}}
{\sqrt{m_{\mu}^2 + (Am_p)^2}} \right]^{-1}.
\eear
$q$ is absolute value of the neutrino momentum, and 
$c_0 c_1 = [1 + ( m_{\mu}/q )^2 ]^{-1/2} $. 
$A$ is the nucleus mass number, while $m_{\mu}$ and $m_p$ are muon and
proton masses, respectively.  All the multipole amplitudes come into
the coefficient,
\bear
\lambda_h = \sum_L \left( {\mid T^{h h}_{L} \mid}^2 + 
{\mid T^{h -h}_{L} \mid}^2 \right).
\eear
Then, we define the weights $\omega_h$;
\bear
\omega_h =  \left( {{1 - 2hc_{0}c_{1}} \over 2} \right) \lambda_h 
\left[ \left( {{1 + c_{0}c_{1}} \over 2} \right) \lambda_- + 
\left( {{1 - c_{0}c_{1}} \over 2} \right) \lambda_+ \right]^{-1}. 
\eear
These weights correspond respectively to the contributions of the
neutrinos with $h = +{1\over 2}$ and $h = -{1 \over 2}$ to the
observables. In the lowest order approximation for the energy-mass
coefficients, eq. (4), (with $m_{\nu}/q \ll 1$) we get 
$\omega_- \simeq 1 $, and $\omega_+ \simeq 0 $.
\\ 
Here we give the recoil polarization formula in the $J_i = 0 \muar J_f
= 1$ transitions;
\bear
\bPf = \sum \limits_{h} \omega_h \bPf(h),
\eear
where
\bear
\bPf(h) =
\left[-2h A_h + \left(A_h - Re B_h \right)( \bPm \cdot \hq ) \right] 
\hq 
+ Re B_h \bPm + 2h Im B_h (\bPm \times \hq).
\eear
The structure functions $A_h$ and $B_h$ are expressed in terms of the
multipole amplitudes ratio;
\bear
\frac{T^{hh}_1}{T^{h-h}_1}  = \sqrt{{1 \over 2}} x_h \eif, 
\eear
\bear
A_h = \frac{2}{2 +  x_h^2}, &\qquad & B_h = \frac{2 x_h \eif }{2 + 
x_h^2}.
\eear
For the longitudinal polarization,
$P_L = {1\over4\pi} \int d\hq <\bPf \cdot \hq >$, we get
\bear
P_L & = & \sum \limits_{h} \omega_h 2h A_h.
\eear
Next we consider angular asymmetry of recoil with respect to the 
direction of 
the muon polarization; 
\bear
{\cal W}(\theta) = 1 + \shel \omega_h 2h \alpha_h \hq \cdot \bPm, 
\eear
where $P_{\mu} \cos \theta = -\hq \cdot \bPm $.
We express the asymmetry coefficients $\alpha_h $ as the functions of
$x_h $ and find their relation to recoil polarization coefficient $A_h
$;
\bear 
\alpha_h = \frac{2 - x_h^2}{2 +  x_h^2} = 2A_h - 1. 
\eear
With the results of [12] for the neutrino polarization, we present 
simple example of the link between $CP$-violation and
massiveness of the Dirac neutrino. We have calculated in accordance 
with the 
definition $<{\bf s}_{\nu}> \equiv Tr\left({\bf 
s}_{\nu}\rho_{f}\right)  
\big/ Tr \overline{\rho}_{f} $; 
\bear 
<{\bf s}_{\nu}>  = \left[a + (c - Reb)(\bPm \cdot \hq)
\right] \hq + Re b \bPm + Imb \hq \times \bPm \ , 
\eear 
where structure functions are given by, 
\bear 
a = \shel h \omega_{h}, & & c = -{1 \over 2}\shel \omega_h
\alpha_h, \\
b &=& c_0c_1 \mnuq {1 \over 2 }T^{- -}_{1} T^{++\ast}_{1}
\left[ \left( {{1 + c_{0}c_{1}} \over 2} \right) \lambda_- + 
\left( {{1 - c_{0}c_{1}} \over 2} \right) \lambda_+ \right]^{-1}. 
\eear 
(Since then, we denote $h \equiv sign \, h$ and $\mu  \equiv sign \, 
\mu $).\\ 
So that, observing in muon capture $<{\bf s}_{\nu } \cdot \hq \times
\bPm > \sim Imb \neq 0$ would indicate both $CP$-violation and massive
neutrino.  For the longitudinal neutrino polarization we have;
\bear 
<{\bf s}_{\nu} \cdot \hq> = a - c \Pm \cos\theta, 
\eear
which gives for $m_{\nu} = 0 $ the recoil asymmetry; 
$ <{\bf s}_{\nu }\cdot \hq > = -{1\over 2} {\cal W}_0(\theta )$.
Averaging this over $\bPm $, we get the helicity for the left-handed 
massless 
neutrino; $<{\bf s}_{\nu}\cdot \hq >_{av} = -{1\over 2}$. 
\section{Discussion and conclusions} 
In the model with massive left-handed neutrino, [12], the existence of
the four independent multipole amplitudes was obtained (doubled in
comparison to the massless case) in the nonrelativistic approximation
of the muon wave function.  As follows from our consideration the
calculation with accurate muon wave function does not change the 
number
of multipole amplitudes, which is still four.  So the form of the
kinematic characteristics of the muon capture expressed by multipole
amplitudes remains the same.  However, the numerical predictions will
change, as the number of nuclear matrix elements in the multipole
amplitudes is larger owing to relativistic muon contribution.  There
appear additional nuclear matrix elements $[1 L L, J]_{hh} $ and $[0 L
L, J_4]_{h -h} $.  Moreover, the structure of other nuclear matrix
elements is also different due to additional terms proportional to
$F(r)$ and $G'(r) $.

Considering nuclear observables, we obtain strict results in terms of
the multipole amplitudes. In addition to standard term, there is
another one with the amplitudes for positive helicity of neutrino. 
This
term multiplies by the weight $\omega_+ $, which is proportional to
$\movq $. In these observables the multipole amplitudes with opposite
neutrino helicities do not interfere.

Additionally , NEMA effects contribute to neutrino Bessel functions in
the multipole amplitudes and very weakly to the phase space volume.
From these, the effect to Bessel function shifts the multipole
amplitudes by a term proportional to $\mvdq $, and doubles its full
value in the capture rate.

On the contrary, the ratio of the multipole amplitudes,
\bear
T^{hh}_1{\big/}T^{h-h}_1,
\eear
weakly depends on NEMA and RLMU effects, which practically disappear 
in
allowed transitions, i.e. for $\pi_i = \pi_f$. In conclusion, such
observables as recoil polarization, recoil asymmetry etc., to a high
degree are independent of these effects.

Qualitative analysis suggests relativistic muon effects larger by
the four orders of magnitude than massive neutrino effects in light
nuclei. In progress, there is our analysis on specific nuclei for
quantitative results. Situation is different in muon capture
reactions with three-body final states, like e.g.;
\bear
(A,Z)_{J_{i}} + \mu^{-} \rightarrow (A-1,Z-1)_{J_{f}} + n + 
\nu_{\mu}, 
\eear
There are experiments performed and planned, 
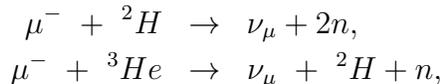
\bear
 \mu^{-}~+~^{2}H  &\rightarrow & \nu_{\mu} + 2n, \nonumber\\
 \mu^{-}~+~^{3}He &\rightarrow & \nu_{\mu}~+~^{2}H + n, \nonumber
 \eear
to measure high energy-momentum transfer to hadrons, which
automatically gives low energy neutrino, $m_{\nu } \big/ E_{\nu }
\rightarrow 1$. Eventual analysis of the data with the massive 
neutrino
theory could be possible.
\vskip 9mm \par
{\bf Acknowledgments:}

The authors are gratefully indebted to Professor F. von Feilitzsch 
for helpful 
discussions.
\vskip 15mm
\section*{References}
\medskip
1. Boehm F and Vogel P 1987 {\em Physics of massive neutrinos} 
(Cambridge: 
University Press)\\
2. Klapdor H V ed 1988 {\em Neutrinos} (Berlin, Heidelberg, New 
York: Springer)\\
3. Dar A and Nussimov S 1991 {\em Part. World} {\bf 2} 117\\
4. Oberauer L and von Feilitzsch F 1992 {\em Rep. Prog. Phys.} {\bf 
55} 1093\\
5. Akhmedov E K, Berezhiani Z G and Senjanovi\'{c} G  1992 {\em Phys. 
Rev. 
Lett.} {\bf 69} 3013\\
6. Berratowicz T, Brannon J, Brazzle R, Cowsik R, Hohenberg C and 
Podosek F 
1992 {\em Phys. Rev. Lett.} {\bf 69} 2341\\
7. Wolfenstein L 1978 {\em Phys. Rev.} {\bf D17} 2369\\
8. Mikheyev S P and Smirnov A Yu 1986 {\em Nuovo Cimento} {\bf 9c} 
17\\
9. Barger V, Phillips R J N and Whisnant K  1992 {\em Phys. Rev. 
Lett.} {\bf 69} 
3135\\
10. Deutsch J P, Lebrun M and Prieels R  1983 {\em Phys. Rev.} {\bf 
D27} 1644\\
11. Kathat C L and Samsonenko N V 1989 {\em Nucl. Phys.} {\bf A491} 
645; 
    {\em J.Phys.} {\bf G15} 1413\\
12. Ciechanowicz S and Popov N 1993 {\em Z.Phys.} {\bf C57} 623\\
13. Oziewicz Z 1970 {\em Partial muon capture}, Ph.D. Thesis, 
University 
of Wroc{\l}aw, ITP \# 208\\
14. Ciechanowicz S and Oziewicz Z 1984 {\em Fortschr. Phys.} {\bf 32} 
61\\
15. Morita M and Fujii A 1960 {\em Phys. Rev. } {\bf 118 } 606 \\
16. Assamagan K et al 1994 {\em Phys Lett.} {\bf B335 } 231 \\
\vskip 3.0 cm
{\bf PACS numbers: 12.15.Pf, 14.60.Gh, 23.40.-s}\\
\medskip
\end{document}